\documentclass[twocolumn]{revtex4}

\usepackage{ragged2e}
\usepackage{amsmath}
\usepackage{graphicx}
\usepackage{bm}
\usepackage[usenames,dvipsnames]{color}

\definecolor{darkblue}{RGB}{0,0,196}

\begin{document}

\title{Multi-Source Thermal Model Describing Transverse Momentum Spectra of
Final-State Particles in High Energy Collisions \vspace{0.5cm}}

\author{Fu-Hu Liu\footnote{E-mail: fuhuliu@163.com;
fuhuliu@sxu.edu.cn}, Jia-Yu Chen, Qiang Zhang}

\affiliation{Institute of Theoretical Physics, State Key
Laboratory of Quantum Optics and Quantum Optics Devices \&
Collaborative Innovation Center of Extreme Optics, Shanxi
University, Taiyuan 030006, China}

\begin{abstract}

\vspace{0.5cm}

\noindent {\bf Abstract:} In this mini review article, the
transverse momentum spectra of final-state particles produced in
high energy hadron-hadron, hadron-nucleus, and nucleus-nucleus
collisions described by the multi-source thermal model at the
quark or parton level is summarized. In the model, the participant
or contributor quarks or partons are considered to contribute
together to the transverse momentum distribution of final-state
particles with different modes of contributions. The concrete mode
of contribution is generally determined by the difference of
azimuthal angles of contributor partons in their emissions.
\\
\\
{\bf Keywords:} Transverse momentum spectra, multi-source thermal
model, Monte Carlo method, convolution method, revised
Tsallis-like function
\\
\\
\\
\end{abstract}

\maketitle

\parindent=15pt

\section{Introduction}

In high energy hadron-hadron, hadron-nucleus, and nucleus-nucleus
collisions, abundant data measured in experiments reflect colorful
mechanisms of particle production and system
evolution~\cite{1,2,3}. As an important issue, the transverse
momentum spectra contain the information of the excitation degree
of emission source, and show the similarity, commonality, and
universality in particle productions~\cite{4,5,6,7,8,9,10,11}. The
multi-source thermal model~\cite{12,13,14,15,16} proposed by us is
successful in describing the distributions of some quantities such
as multiplicities, isotopic cross-sections, (pseudo)rapidities,
transverse energies, azimuthal angles, transverse momenta, etc. In
the model, the nucleons or nucleon clusters were regarded as the
multi-source and the Boltzmann-Gibbs statistics was used in
describing a given source.

The multi-source thermal model was proposed according to the
single-, two-, and three-fireball
models~\cite{17,18,19,20,21,22,23,24}, as well as the multi-source
ideal-gas model or the cylinder model~\cite{25,26,27,28}.
Recently, the Boltzmann-Gibbs statistics used in the model was
replaced by the Tsallis statistics, and the multi-source of
fireballs (nucleons or nucleon clusters) was replaced by the
multi-source of participant or contributor quarks or
partons~\cite{29,30,31}. Here, the single component distribution
from the Boltzmann-Gibbs statistics is not enough to fit the
transverse momentum spectra. Two- or three-component distribution
is needed, which results in the temperature fluctuations and is
covered by the Tsallis distribution with less
parameters~\cite{31a}.

It should be noted that in the multi-source thermal model the
sources changed from fireballs to participant partons means that
the smaller contributor units at the deeper level are used. This
is an important progress or improvement in the viewpoint of the
model. A fireball may contain lots of partons, and the partons are
the underlying units of collisions. The latest version of the
model was tested firstly by the transverse momentum spectra of
final-state particles~\cite{29,30,31}. Some quantities such as the
temperature of parton source and the average transverse flow
velocity of partons can be extracted from describing the
transverse momentum spectra. In view of the latest progress of the
model, it is necessary to review and summarize it in some way.

The multi-source thermal model is a static thermodynamical and
statistical model. Although the dynamical evolution process of the
interacting system cannot be described by the model, some useful
quantities can be extracted from the comparisons of the model with
experimental data. The dependences of the concerned quantities on
collision energy, event centrality, system size, particle
rapidity, particle mass, and quark mass can be obtained. The
sudden changes of these dependences are expected to relate to the
formation of quark-gluon plasma (QGP) or quark
matter~\cite{31,31a0,31a1,31a2,31a3,31a4,31a5}. It is believed
that QGP was produced in a hot and dense environment formed in the
experiments at the relativistic heavy ion collider
(RHIC)~\cite{31a7,31a8,31a9,31a10,31a11,31a12} and the large
hadron collider (LHC)~\cite{31a13,31a14,31a15,31a16}. As the
strongly interacting partonic medium formed at the RHIC and LHC,
QGP was predicted by the quantum chromodynamics (QCD) theory which
describes the strong
interactions~\cite{31a17,31a18,31a19,31a20,31a21}.

This mini review article will summarize the method for describing
the transverse momentum spectra of final-state particles produced
in high energy collisions in the framework of multi-source thermal
model at the parton level. The contributions of the contributor
partons are considered in different ways where different azimuthal
differences are used. The azimuthal difference may be various
values in $[0,2\pi]$ in general, or in particular 0 or $\pi$ if
the contributions of two partons are parallel, or $\pi/2$ if the
contributions of two partons are perpendicular. Although the
azimuthal difference may be particular value, the azimuthal angles
are independent and there is no sorting for them.

The rest of this article is structured as follows. The physics
picture and formalism expression of the multi-source thermal model
at the parton level are described in Section 2. Implementation and
discussion are given in Section 3. In Section 4, the summary and
conclusion of this article are given.

\begin{figure*}[htbp]
\begin{center}\vspace{.0cm}\hspace{-.0cm}
\includegraphics[width=8.0cm]{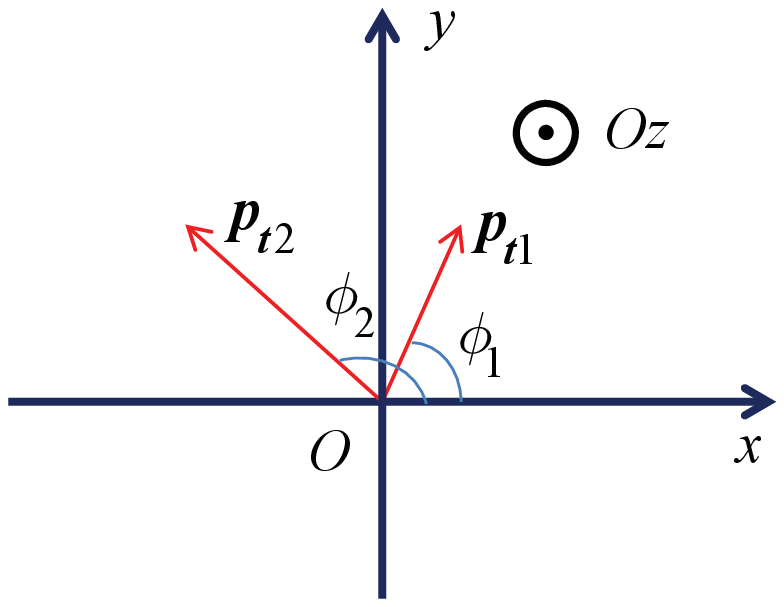}
\end{center}\vspace{0.25cm}
\justifying\noindent {\small Fig. 1. The right-handed Cartesian
coordinate system \(O\)-\(xyz\). The beam direction which is along
the \(Oz\) axis points from the inside to the outside. The
reaction plane is the plane \(xOz\), and the transverse plane is
the plane \(xOy\) which is perpendicular to the beam direction. In
the figure, \(\phi_1\) (\(\phi_2\)) is the angle of vector \(\bm
p_{t1}\) (\(\bm p_{t2}\)) measured with respect to the \(Ox\) axis
in the plane \(xOy\).}
\end{figure*}

\section{Picture and formalism}

In high energy hadron-hadron, hadron-nucleus, and nucleus-nucleus
collisions, many final-state particles are produced in collision
process and measured in experiments. Meanwhile, a few fragments
which are nucleons or nucleon clusters from the spectator
fragmentation are produced in the final state in hadron-nucleus
and nucleus-nucleus collisions. In collisions at very high
energies, a few jets are produced, which consists of many
particles. In most cases, final-state particles are main products
in high energy collisions.

To describe the production of final-state particles, it is natural
that a single-fireball is assumed to form in the collisions of
projectile and target hadrons (or nuclei) at a few GeV which is
not too high energy. In the rest frame of the fireball, one may
assume that the particles are emitted isotropically, as discussed
in the multi-source thermal model~\cite{12,13,14,15,16}. However,
the particles are anisotropic in experiments. Then, the
single-fireball is needed to extend to a two-fireball~\cite{17,18}
in which one is from the projectile hadron (or nucleus) and the
other one is from the target hadron (or nucleus), or a
three-fireball~\cite{19,20,21,22,23,24} which consists of the
projectile, central, and target fireballs. Although the particles
are isotropic in the rest frame of each fireball, the experimental
spectra can be anisotropic due to the motion of the fireball.

Further, the three-fireball is extended to a thermalized cylinder
or fire-cylinder~\cite{31aa1,31aa2,31aa3,31aa4} which is formed
due to the penetrations of projectile and target hadrons (or
nuclei) in the collisions at higher energy (dozens of GeV and
TeV). The single-cylinder can be extended to a
two-cylinder~\cite{26,27} in which one is from the projectile
hadron (or nucleus) and the other one is from the target hadron
(or nucleus). The two cylinders may overlap or separate each
other. In the rapidity space, the emission points with the same
rapidity in the cylinder(s) consist of a large emission source. In
the rest frame of the considered large emission source, the
particles are assumed to emit isotropically.

Generally, a given particle is produced from the interactions of
two or three contributor partons. Concretely, a meson (baryon) is
produced from the interactions of two (three) constituent or
contributor quarks, while a lepton is produced from the
interactions of two contributor quarks or gluons. Here, the
additive quark model~\cite{31b,31c,31d,31e,31f} is considered for
part case, in which the meson (baryon) consists of two (three)
constituent quarks within the model, but these are not numbers of
quarks producing meson (baryon) via their (quarks) interactions
which are additionally considered in the multi-source thermal
model~\cite{12,13,14,15,16}. In most cases, more partons may take
part in the interactions. However, only two or three partons take
part in the main role in the production of a given particle. Of
course, for a tetraquark or pentaquark state, one naturally
considers four or five constituent quarks. As an approximate
treatment, two contributor heavy quarks may also produce a
multiquark state or an arbitrary jet.

Let \(p_T\) and \(p_{t1}\) (\(p_{t2}\)) denote the transverse
momentum of given particle and the contribution amount of the
first (second) parton to $p_T$ respectively. The probability
density function obeyed by \(p_T\) and \(p_{t1}\) (\(p_{t2}\)) are
\(f(p_T)\) and \(f_1(p_{t1})\) (\(f_2(p_{t2})\)) respectively,
where the variables such as the temperature parameter \(T\) and
entropy index \(q\) in the Tsallis statistics are not listed in
the functions for convenience. One may study the relation between
\(p_T\) and \(p_{t1}\) (\(p_{t2}\)), as well as \(f(p_T)\) and
\(f_1(p_{t1})\) (\(f_2(p_{t2})\)) according to the difference
between the azimuthal angle \(\phi_1\) of the first parton and the
azimuthal angle \(\phi_2\) of the second parton in the emission.

We would like to explain the azimuthal angle in the right-handed
Cartesian coordinate system \(O\)-\(xyz\) in detail. For clarity,
Figure 1 shows the scheme of kinematic variables in the transverse
plane \(xOy\), where the beam direction which is along the \(Oz\)
axis points from the inside to the outside and the reaction plane
is \(xOz\). Here, \(\phi_1\) (\(\phi_2\)) is the angle of vector
\(\bm p_{t1}\) (\(\bm p_{t2}\)) measured with respect to the
\(Ox\) axis in the transverse plane \(xOy\) which is perpendicular
to the beam direction \(Oz\) axis.

In the following text, a general case and two particular cases are
discussed in subsections i)--iii) successively. Then, the
connection of \(p_T\) to the rapidity \(y\) and pseudorapidity
\(\eta\) is discussed in subsection iv). For each issue, the basic
method and formalism are presented.
\\

\noindent \emph{i) General case: various azimuths}

For any difference between \(\phi_1\) and \(\phi_2\), the analytic
relation between \(f(p_T)\) and \(f_1(p_{t1})\) (\(f_2(p_{t2})\))
is hard to obtain. Instead, one may use the Monte carlo method to
perform the calculations. Based on \(f_1(p_{t1})\) and
\(f_2(p_{t2})\), one may obtain \(p_{t1}\) and \(p_{t2}\) firstly.
In fact, in the Monte Carlo method, let \(r_{1,2,3,4}\) denote
random numbers distributed evenly in \([0,1]\). One may extract
\(p_{t1}\) and \(p_{t2}\) according to
\begin{align}
\int_0^{p_{t1}} f_{1}(p'_{t1})dp'_{t1} <r_{1} \leq
\int_0^{p_{t1}+\delta p_{t1}} f_{1}(p'_{t1})dp'_{t1}, \\
\int_0^{p_{t2}} f_{2}(p'_{t2})dp'_{t2} <r_{2} \leq
\int_0^{p_{t2}+\delta p_{t2}} f_{2}(p'_{t2})dp'_{t2}.
\end{align}
Meanwhile, one may obtain \(\phi_1\) and \(\phi_2\) due to the
assumption of isotropy in the source rest frame. That is
\begin{align}
\phi_1&=2\pi r_3, \\
\phi_2&=2\pi r_4
\end{align}
in the Monte Carlo method, where \(\phi_1\) (\(\phi_2\))
distributes evenly in \([0,2\pi]\).

Because isotropic azimuth obeys the uniform distribution,
\(f_{\phi}(\phi)=1/(2\pi)\), in \([0,2\pi]\) in the transverse
plane, the expressions of Eqs. (3) and (4) are natural. It should
be noted that both \(\phi_1\) and \(\phi_2\) are the independent
random numbers distributed evenly in \([0,2\pi]\) due to the fact
that they come from the independent random numbers \(r_3\) and
\(r_4\) distributed evenly in \([0,1]\) respectively. There is no
sorting for \(\phi_1\) and \(\phi_2\) when they are performed
through Eqs. (3) and (4) respectively. Different lower footmarks
are used for the two azimuths due to different values. In fact,
Eqs. (3) and (4) indeed describe isotropic azimuth respectively,
if \(\int_0^{\phi_1}f_{\phi}(\phi')d\phi'=r_3\) and
\(\int_0^{\phi_2}f_{\phi}(\phi')d\phi'=r_4\) are solved. In the
Monte Carlo calculations in the multi-source thermal
model~\cite{12,13,14,15,16}, \(\phi\), but not \(\cos\phi\), is
used because both \(\cos\phi\) and \(\sin\phi\) can be used more
easily.

The two components \(p_{x}\) and \(p_{y}\), as well as \(p_T\)
itself can be given by
\begin{align}
p_x &= p_{t1}\cos\phi_1+p_{t2}\cos\phi_2, \\
p_y &= p_{t1}\sin\phi_1+p_{t2}\sin\phi_2, \\
p_T &=\sqrt{p_x^2+p_y^2} \nonumber\\
&=\sqrt{p_{t1}^2+p_{t2}^2+2p_{t1}p_{t2}\cos|\phi_1-\phi_2|}.
\end{align}
Then, the probability density function, \((1/N)dN/dp_T\), of
\(p_T\) can be obtained by the statistics, where \(N\) denotes the
number of particles. For the frequently-used experimental
spectrum, \((1/2\pi p_T)d^2N/dp_Tdy\), and other forms, the
statistical method is also available to obtain the forms
correspondingly.

The above method can be easily extended to the case of three
contributor partons. The quantities \(p_{t3}\), \(f_3(p_{t3})\),
and \(\phi_3\) related to the third component can be obtained by
the same method. For the case of more partons, the method is also
applicable. What one does is considering the third or more
components in the expression of \(p_x\) and \(p_y\). In fact,
including the third parton, one has more equations
\begin{align}
\int_0^{p_{t3}} f_{3}(p'_{t3})dp'_{t3}< & r_{5} \leq
\int_0^{p_{t3}+\delta p_{t3}} f_{3}(p'_{t3})dp'_{t3}, \\
& \phi_3=2\pi r_6,
\end{align}
where \(r_{5,6}\) are random numbers distributed evenly in
\([0,1]\) due to the requirement in the Monte Carlo method. The
two components \(p_{x}\) and \(p_{x}\) are improved by
\begin{align}
p_x &= p_{t1}\cos\phi_1+p_{t2}\cos\phi_2+p_{t3}\cos\phi_3, \\
p_y &= p_{t1}\sin\phi_1+p_{t2}\sin\phi_2+p_{t3}\sin\phi_3,
\end{align}
in which one more item is added. One has
\begin{align}
p_T=&\sqrt{p_x^2+p_y^2}\nonumber\\
=&\big(p_{t1}^{2}+p_{t2}^{2}+p_{t3}^{2}+2p_{t1}p_{t2}\cos|\phi_{1}-\phi_{2}|\nonumber\\
&+2p_{t1}p_{t3}\cos|\phi_{1}-\phi_{3}|+2p_{t2}p_{t3}\cos|\phi_{2}-\phi_{3}|\big)^{1/2}.
\end{align}
The case of more partons can be conveniently considered by the
frequently-used method of vector synthesis.
\\

\begin{figure*}[htbp]
\begin{center}\vspace{1.0cm}\hspace{0.5cm}
\includegraphics[width=10.0cm]{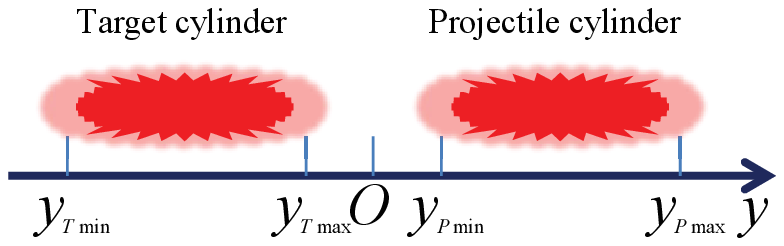}
\end{center}\vspace{0.25cm}
\justifying\noindent {\small Fig. 2. The relative positions of
$y_{P\min}$, $y_{P\max}$, $y_{T\min}$, and $y_{T\max}$ in the
rapidity space. The projectile cylinder is assumed to appear on
the right side, while the target cylinder appears on the left
side.}
\end{figure*}

\noindent \emph{ii) Particular case: parallel transverse momenta}

For a particular case of \(\phi_1-\phi_2=0\), one has
\(p_T=p_{t1}+p_{t2}\). The Monte Carlo method discussed above is
naturally applicable. In addition, the analytic relation between
\(f(p_T)\) and \(f_1(p_{t1})\) (\(f_2(p_{t2})\)) is easy to
obtain~\cite{29,30}. In fact, \(f(p_T)\) is the convolution of
\(f_1(p_{t1})\) and \(f_2(p_{t2})\). One has
\begin{align}
f(p_{T})&=\int_0^{p_{T}} f_1(p_{t1})f_2(p_{T}-p_{t1})dp_{t1} \nonumber\\
&=\int_0^{p_{T}} f_2(p_{t2})f_1(p_{T}-p_{t2})dp_{t2}.
\end{align}

If the case of three contributor partons with the same azimuthal
angle is considered, one has \(p_T=p_{t1}+p_{t2}+p_{t3}\). Except
the Monte Carlo method, the convolution method is also useable.
One has the convolution of \(f_1(p_{t1})\) and \(f_2(p_{t2})\) to
be
\begin{align}
f_{12}(p_{t12})&=\int_0^{p_{t12}} f_1(p_{t1}) f_2(p_{t12}-p_{t1})
dp_{t1} \nonumber\\
&=\int_0^{p_{t12}} f_2(p_{t2}) f_1(p_{t12}-p_{t2}) dp_{t2}.
\end{align}
The convolution of \(f_{12}(p_{t12})\) and the third function
\(f_3(p_{t3})\) is
\begin{align}
f(p_{T})&=\int_0^{p_{T}} f_{12}(p_{t12}) f_3(p_{T}-p_{t12})
dp_{t12}
\nonumber\\
&=\int_0^{p_{T}} f_3(p_{t3}) f_{12}(p_{T}-p_{t3}) dp_{t3}.
\end{align}
The case of more partons can be considered by the convolution
method step by step~\cite{29,30}.
\\

\noindent \emph{iii) Particular case: vertical transverse momenta}

For a particular case of \(|\phi_1-\phi_2|=\pi/2\), one has
\(p_T=\sqrt{p^2_{t1}+p^2_{t2}}\). The Monte Carlo method discussed
above is naturally applicable. In addition, the analytic relation
between \(f(p_T)\) and \(f_1(p_{t1})\) (\(f_2(p_{t2})\)) is easy
to obtain~\cite{31,32,32a}. Let \(\phi\) (or \(\pi/2-\phi\))
denote the azimuthal angle of the vector \(\bm{p_T}\) relative to
the vector \(\bm{p_{t1}}\) (or the vector \(\bm{p_{t2}}\)). One
has the united probability density function of \(p_T\) and
\(\phi\) to be
\begin{align}
f_{p_T,\phi}(p_T,\phi) &= p_Tf_{1,2}(p_{t1},p_{t2}) \nonumber\\
&= p_Tf_1(p_{t1})f_2(p_{t2}) \nonumber\\
&= p_Tf_1(p_T\cos\phi)f_2(p_T\sin\phi),
\end{align}
where \(f_{1,2}(p_{t1},p_{t2})\) is the united probability density
function of \(p_{t1}\) and \(p_{t2}\). By integrating \(\phi\),
one has
\begin{align}
f(p_T)&=\int_{0}^{2\pi}f_{p_T,\phi}(p_T,\phi)d\phi \nonumber\\
&=p_T\int_{0}^{2\pi}f_1(p_T\cos\phi)f_2(p_T\sin\phi)d\phi.
\end{align}

For the case of three partons, if the synthesis of \(\bm{p_{t1}}\)
and \(\bm{p_{t2}}\) is coincidentally perpendicular to the vector
\(\bm{p_{t3}}\), the method of united probability density function
is still applicable. Let the vector \(\bm{p_{t12}}\) denote the
synthesis of \(\bm{p_{t1}}\) and \(\bm{p_{t2}}\), \(\phi_{12}\)
denote the azimuthal angle of \(\bm{p_{t12}}\) relative to
\(\bm{p_{t1}}\), and \(\phi'\) denote the azimuthal angle of
\(\bm{p_T}\) relative to \(\bm{p_{t12}}\). The united probability
density function of \(p_{t12}\) and \(\phi_{12}\) is
\begin{align}
f_{p_{t12},\phi_{12}}(p_{t12},\phi_{12}) &= p_{t12}f_{1,2}(p_{t1},p_{t2}) \nonumber\\
&= p_{t12}f_1(p_{t1})f_2(p_{t2}) \nonumber\\
&= p_{t12}f_1(p_T\cos\phi_{12})f_2(p_T\sin\phi_{12}).
\end{align}
By integrating \(\phi_{12}\), one has
\begin{align}
f_{12}(p_{t12})&=\int_{0}^{2\pi}f_{p_{t12},\phi_{12}}(p_{t12},\phi_{12})d\phi_{12} \nonumber\\
&=p_{t12}\int_{0}^{2\pi}f_1(p_{t12}\cos\phi_{12})f_2(p_{t12}\sin\phi_{12})d\phi_{12}.
\end{align}

The united probability density function of \(p_T\) and \(\phi'\)
is
\begin{align}
f_{p_T,\phi'}(p_T,\phi') &= p_Tf_{12,3}(p_{t12},p_{t3}) \nonumber\\
&= p_Tf_{12}(p_{t12})f_3(p_{t3}) \nonumber\\
&= p_Tf_{12}(p_T\cos\phi')f_3(p_T\sin\phi'),
\end{align}
where \(f_{12,3}(p_{t12},p_{t3})\) is the united probability
density function of \(p_{t12}\) and \(p_{t3}\). By integrating
\(\phi'\), one has
\begin{align}
f(p_T)&=\int_{0}^{2\pi}f_{p_T,\phi'}(p_T,\phi')d\phi' \nonumber\\
&=p_T\int_{0}^{2\pi}f_{12}(p_T\cos\phi')f_3(p_T\sin\phi')d\phi'.
\end{align}
The case of more partons can be considered by the method of united
probability density function step by step~\cite{31,32,32a}.
\\

\noindent \emph{iv) Connection of \(p_T\) to (pseudo)rapidity}

In the rapidity space, the projectile cylinder is assumed to stay
in the rapidity range \([y_{P\min},y_{P\max}]\), and the target
cylinder stays in the rapidity range \([y_{T\min},y_{T\max}]\).
Figure 2 gives the relative positions of the four rapidities. It
assumes that the projectile (target) comes from the left (right)
side and the projectile (target) cylinder appears on the right
(left) side. If \(y_{P\min}<y_{T\max}\), the two cylinders
overlap. If \(y_{P\min}>y_{T\max}\), there is a gap between the
two cylinders. If \(y_{P\min}=y_{T\max}\), the two cylinders are
connected into one. For symmetrical collisions, one has the
relations: \(y_{P\max}-y_{P\min}=y_{T\max}-y_{T\min}\),
\(y_{P\min}=-y_{T\max}\), \(y_{P\max}=-y_{T\min}\). These
relations reduce the number of parameters.

In the Monte carlo method, let \(R_{1,2}\) denote random numbers
distributed evenly in \([0,1]\). One has the rapidity \(y_x\) of
the emission source distributed evenly in
\([y_{P\min},y_{P\max}]\) and \([y_{T\min},y_{T\max}]\) to be
\begin{align}
y_x&=y_{P\min}+(y_{P\max}-y_{P\min})R_1,\\
y_x&=y_{T\min}+(y_{T\max}-y_{T\min})R_2,
\end{align}
respectively.

In the source rest frame, in the case of isotropic emission, the
probability density function, \(f_{\theta'}(\theta')\), of the
emission angle, \(\theta'\), of the considered particle obeys the
half sine function, \((1/2)\sin\theta'\), which results in
\begin{align}
\theta'=2\arcsin\sqrt{R_3},
\end{align}
where \(R_{3}\) denotes random number distributed evenly in
\([0,1]\). Then, one has the momentum \(p'\), longitudinal
momentum \(p'_z\), and energy \(E'\) to be
\begin{align}
p'&= p_T\csc\theta',\\
p'_z&= p_T\cot\theta',\\
E'&= \sqrt{p'^2+m_0^2},
\end{align}
where \(m_0\) is the rest mass of the considered particle.

The rapidity \(y'\) in the source rest frame is
\begin{align}
y'= \frac{1}{2}\ln\left(\frac{E'+p'_z}{E'-p'_z}\right).
\end{align}
The rapidity \(y\) measured in experiments is
\begin{align}
y=y'+y_x.
\end{align}
The rapidity distribution, \((1/N)dN/dy\), can be obtained by the
statistics.

In terms of the pseudorapidity, \(\eta\), measured in experiments,
one needs the longitudinal momentum
\begin{align}
p_z=\sqrt{p_T^2+m_0^2}\sinh y.
\end{align}
Then, the emission angle
\begin{align}
\theta=\arctan\frac{p_T}{p_z}
\end{align}
and the pseudorapidity
\begin{align}
\eta=-\ln\tan\frac{\theta}{2}.
\end{align}
The pseudorapidity distribution, \((1/N)dN/d\eta\), can be
obtained by the statistics. Here, \(\eta\) and \(y\), and their
distributions, are obtained respectively.

In particular, if the analytical expression, \(f_{y'}(y')\), of
the probability density function for \(y'\) in the source rest
frame is available, one has the analytical expression,
\(f_{y}(y)\), of the probability density function for \(y\) in
experiments to be
\begin{align}
f_{y}(y)=& \frac{k}{y_{P\max}-y_{P\min}}
\int_{y_{P\min}}^{y_{P\max}}f_{y'}(y-y_x)dy_x \nonumber\\
&+ \frac{1-k}{y_{T\max}-y_{T\min}}
\int_{y_{T\min}}^{y_{T\max}}f_{y'}(y-y_x)dy_x,
\end{align}
where \(k\) (\(1-k\)) denotes the contribution fraction of the
projectile (target) cylinder.

\section{Implementation and discussion}

In the general case and the two particular cases discussed above,
one needs firstly to choose \(f_1(p_{t1})\), \(f_2(p_{t2})\), and
\(f_3(p_{t3})\) for contributor partons. The three functions
should be the same in form with the same or different parameter
values. To find a suitable function, one has tried many attempts.
Finally, one finds that the Tsallis function is a possible
candidate, though it has different forms in the literature,
including in high energy physics~\cite{32b,32c,32d,32e,32f}. In
particular, the revised Tsallis-like function is a suitable choice
according to our recent attempts~\cite{29,30}. Then, one has
\begin{align}
f_i(p_{ti}) = C_i p_{ti}^{a_0} \left[ 1-\frac{1-q}{T} \left(
m_{ti}-m_{0i}\right) \right]^{q/(1-q)},
\end{align}
where the subscript \(i\) is for the \(i\)-th contributor parton,
\(m_{ti}=\sqrt{p_{ti}^2+m_{0i}^2}\) is the transverse mass,
\(m_{0i}\) is the empirical constituent mass, \(T\) is the
effective temperature, \(q\) is the entropy index, $a_0$ is the
revised index, and \(C_i\) is the normalization constant. The
power index \(q/(1-q)\) is used to cater to the consistency of
thermodynamics from the probabilities of microstates and the
maximum entropy principle.

In Eq. (34), the entropy index \(q\) describes the departure
degree of the system form the equilibrium, or the degree of
non-equilibrium of the system. Generally, \(q=1\) corresponds to
the equilibrium, \(1<q<1.25\) means an approximate equilibrium. As
an insensitive quantity, \(q\) is not too large even at very high
energy, which means the approximate equilibrium of the system.
Empirically, for both the quarks and gluons, \(m_{0i}\) is
regarded as the constituent masses of quarks, but not other mass
such as the bare or the effective quark mass. For light (heavy)
particles, \(m_{0i}\) is taken to be the constituent masses of
light (heavy) quarks. For various jets, \(m_{0i}\) is taken to be
the constituent masses of heavy quarks. The specific quarks (with
different \(m_{0i}\)) depend on the types of particles and jets,
which is used in our recent work~\cite{29,30,31,45}.

\begin{figure*}[htbp]
\begin{center}
\includegraphics[width=12.0cm]{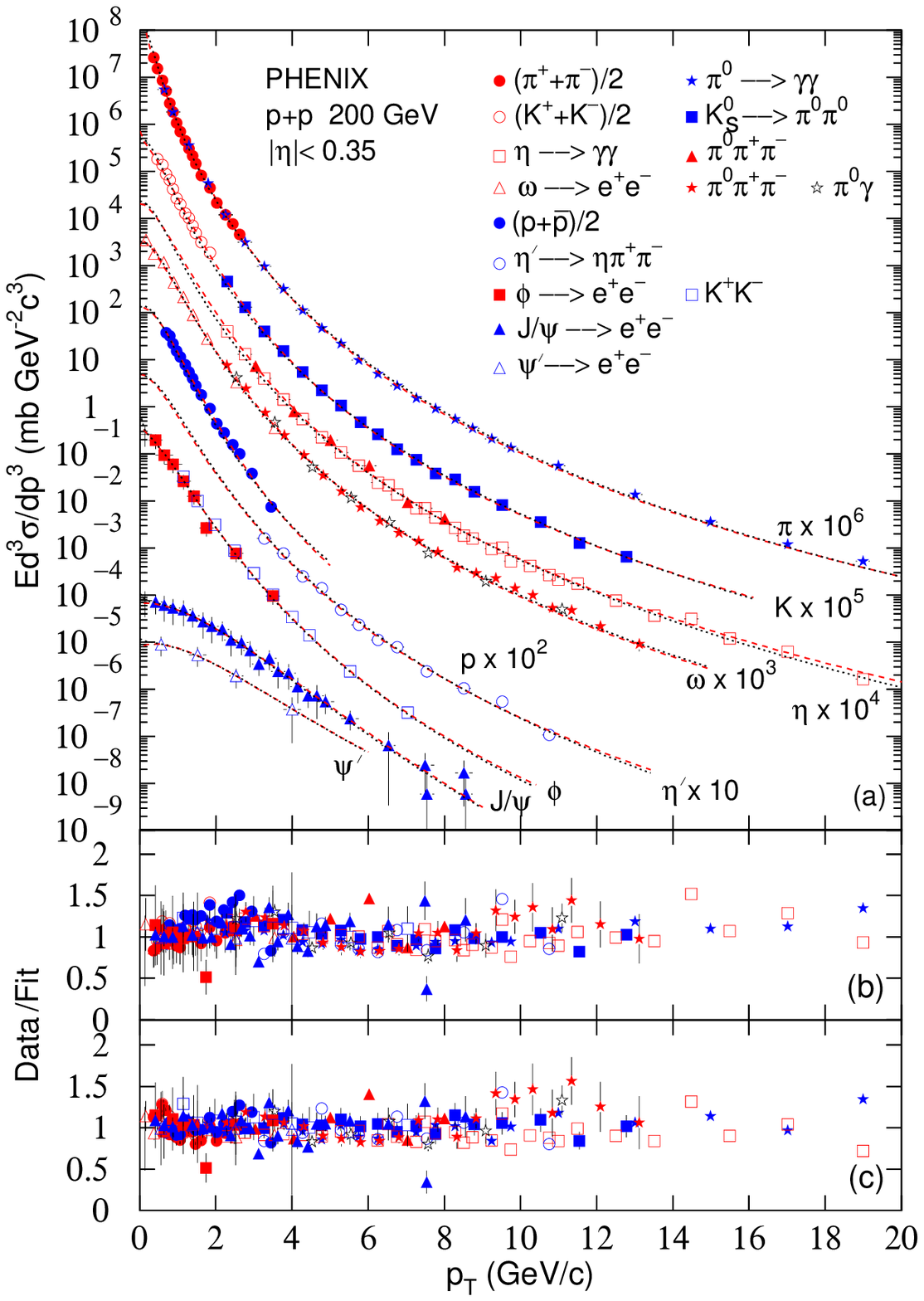}
\end{center}
\justifying\noindent {\small Fig. 3. (a) The invariant
cross-sections of various hadrons with given combinations and
decay channels produced in p+p collisions at 200 GeV. The symbols
represent the experimental data measured by the PHENIX
Collaboration~\cite{46a}, and the dotted and dashed curves are the
fitted results by using the revised Tsallis distribution [with the
power index \(1/(1-q)\) in Eq. (34)] and the convolution,
respectively~\cite{29}. (b) The ratio of data to fit obtained from
the dotted curves. (c) The ratio of data to fit obtained from the
dashed curves. Detailed information and related parameters can be
found in ref.~\cite{29} from where the figure is cited.}
\end{figure*}

The above revised Tsallis-like function is possibly to be revised
again for different cases. For example, for the particular case of
parallel transverse momenta, it is suitable. For the particular
case of vertical transverse momenta~\cite{31,32}, one has another
revision
\begin{align}
f_i(p_{ti}) = C_i m_{ti}^{a_0} \left[ 1-\frac{1-q}{T} \left(
m_{ti}-m_{0i}\right) \right]^{q/(1-q)}.
\end{align}
The two revisions are empirical expressions. The first revision
can be used in the second particular case approximately.
Meanwhile, the second revision can be used in the first particular
case approximately.

The above concrete expressions of \(f_i(p_{ti})\) are for
mid-rapidity (mid-\(y_i\approx 0\)) only. Meanwhile, the chemical
potential (\(\mu_i\)) is not included. To include non-mid-\(y_i\)
and non-zero \(\mu_i\), one may use \(m_{ti}\cosh
y_i-\mu_i-m_{0i}\) to replace \(m_{ti}-m_{0i}\) expediently, and
perform an integral for \(y_i\) from the minimum to maximum
\(y_i\). If the range from the minimum to maximum \(y_i\) does not
cover the mid-\(y_i\), one may shift the range to cover the
mid-\(y_i\) simply. This performance is to exclude the
contribution of directed and longitudinal motion of the emission
source from the temperature parameter.

The parameter \(T\) is called the effective temperature, but not
the (physical) temperature, due to the fact that the contribution
of flow  effect is not excluded. To dissociate the contributions
of thermal motion and flow effect related to \(i\)-th contributor
parton, one may use an alternative method in which the intercept
in the linear relation of \(T\) versus \(m_{0i}\) is regarded as
the kinetic freeze-out temperature, and the slope in the linear
relation of average \(p_{ti}\) (\(\langle p_{ti}\rangle\)) versus
average moving mass (\(\overline{m}_i\)) or average energy
(\(\overline{E}_i\)) in the source rest frame is regarded as the
average transverse flow velocity~\cite{33a,33b,33c,33d,33,34}.

The above alternative method of intercept-slope works well at the
particle level. It does not work well at the parton level due to
limited type and undefined mass of partons. For example, in many
cases, one has only one or two types of partons available in the
analysis, and the masses of up and down quarks are almost the
same. In addition, the mass of gluon has no strict definition in
the analysis, though one may regard it as the constituent mass of
light quarks approximately if needed. These limited type and
undefined mass of partons result in the application of the
alternative method of intercept-slope not to be on the right way.

To obtain the kinetic freeze-out temperature \(T_0\) and the
average transverse flow velocity \(\langle\beta_t\rangle\) at the
parton level, one may perform a Lorentz-like transformation for
\(p_{ti}\) and \(m_{ti}\) in the concrete expressions of
\(f_i(p_{ti})\). For clarity, \(p_{ti}\), \(m_{ti}\), and
\(f_i(p_{ti})\) in Eqs. (34) and (35) are substituted by
\(p'_{ti}\), \(m'_{ti}\), and \(f_{i'}(p'_{ti})\), respectively.
One has the transformations
\begin{align}
|p'_{ti}|
=\langle\gamma_t\rangle|p_{ti}-m_{ti}\langle\beta_t\rangle|,\\
m'_{ti}
=\langle\gamma_t\rangle(m_{ti}-p_{ti}\langle\beta_t\rangle),
\end{align}
where
\(\langle\gamma_t\rangle=1/\sqrt{1-\langle\beta_t\rangle^2}\)~\cite{35,36,37,38}
is the Lorentz-like factor. It should be noted that the
Lorentz-like, but not the Lorentz, transformation or factor is
called due to the fact that \(\langle\beta_t\rangle\) is used, but
not \(\beta_t\). The absolute value
\(|p_{ti}-m_{ti}\langle\beta_t\rangle|\) is used due to
\(p'_{ti}\) being positive and
\(p_{ti}-m_{ti}\langle\beta_t\rangle\) being possibly negative in
low-\(p_{ti}\) region. After the conversion, one has
\begin{align}
f_i(p_{ti}) &= f_{i'}(p'_{ti})\left|\frac{dp'_{ti}}{dp_{ti}}\right| \nonumber\\
&= f_{i'}(\langle\gamma_t\rangle|p_{ti}
-m_{ti}\langle\beta_t\rangle|)\langle\gamma_t\rangle
\frac{m_{ti}-p_{ti}\langle\beta_t\rangle}{m_{ti}}
\end{align}
because of the probability conservation, where $f_{i'}(p'_{ti})$
is given by Eqs. (34) and (35) due to the substitution before the
conversion. After the conversion, the parameter $T$ in the
concrete expressions of \(f_i(p_{ti})\) is the kinetic freeze-out
temperature $T_0$.

The Monte Carlo method is suitable to the general case where the
difference between azimuthal angles are various. However, the
calculated results have no value or have large fluctuations in
high-\(p_T\) region, even the total number of simulated particles
is very large. One needs to improve the method of simulated
calculation, for example, using the piecewise simulation for the
low- and high-\(p_T\) regions respectively. By contrary, the
analytical method suited to the parallel or perpendicular
situation describes the spectra in whole \(p_T\) region smoothly.

Our recent studies~\cite{29,30,31,45} show that the convolution
method for the parallel case is more easier to fit the wide
\(p_T\) spectra. A two-component function is needed for the wide
\(p_T\) spectra if the method of united probability density
function for the perpendicular case is used. The Monte Carlo
method for the general case seems more reasonable, though more
computing resources are needed. According to the fitting
experience, to fit the spectra in wide \(p_T\) range, the
convolution method for the parallel case is more convenient.

As an example of the application of the multi-source thermal
model, the \(p_T\) spectra (the invariant cross-section),
\(Ed^3\sigma/dp^3\), of various hadrons with given combinations
and decay channels produced in proton-proton (p+p) collisions at
center-of-mass energy of 200 GeV is displayed in Figure 3 which is
cited from ref.~\cite{29}. In panel (a), the symbols represent the
experimental data measured by the PHENIX Collaboration~\cite{46a},
and the dotted and dashed curves are the fitted results by using
the revised Tsallis distribution [with the power index \(1/(1-q)\)
which has a slight difference from Eq. (34)] and the convolution,
respectively, where \(\langle\beta_t\rangle\) has not yet been
introduced~\cite{29}. In panels (b) and (c), the ratios of data to
fit are presented corresponding to the dotted and dashed curves
respectively. Detailed information on Figure 3 and the related
parameters can be found in ref.~\cite{29}. More figures can be
found in refs.~\cite{29,30,31,45}, which studies various particles
and jets produced in different collisions over an energy range
from a few GeV to above 10 TeV.

In the above discussions, the particles, collisions, and energies
are not distinguished deliberately. In fact, not only for baryons
but also for leptons, one may use the same idea and formalism to
fit their \(p_T\) spectra in wide range in different collisions at
different energies~\cite{29,30,31,45}. In particular, in most
cases, the number of participant partons is defined by two. This
is due to one projectile parton and one target parton being main
participants. Even for the \(p_T\) spectra of various jets, one
may use the convolution of two revised Tsallis-like functions to
fit them~\cite{45}. The third participant parton maybe is needed
to revise the result of two participant partons. From small system
such as hadron-hadron and hadron-nucleus collisions to large
system such as nucleus-nucleus collisions, the idea and formalism
are the same. This sameness is a reflection of the similarity,
commonality, and universality existed in high energy
collisions~\cite{4,5,6,7,8,9,10,11}. This enlightens that the
contributions of contributor partons for different particles in
different collisions are considered.

In addition, not only for the spectra in low-\(p_T\) region which
is contributed by the soft excitation process, but also for the
spectra in high-\(p_T\) region which is contributed by the hard
scattering process, one has uniformly used the same idea and
formalism. In the framework of multi-source thermal model at the
parton level, both the processes are considered due to the
contributions of contributor partons. There is no obvious
difference for the two processes, but the violent degree. This
sameness is also a reflection of the similarity, commonality, and
universality existed in high energy
collisions~\cite{4,5,6,7,8,9,10,11}.

Before summary and conclusion, it should be emphasized that the
parameter \(\langle\beta_t\rangle\) is the average transverse flow
velocity at the parton level. Although both \(p_T\) and \(y\) can
be deformed by the presence of collective flow, a suitable
description for the spectrum of particles naturally includes the
influence of collective flow. As a constant value for given
spectrum, \(\langle\beta_t\rangle\) is independent of \(p_T\).
However, \(\langle\beta_t\rangle\) depends on \(y\) due to the
fact that the spectrum depends on \(y\). The introduction of
\(\langle\beta_t\rangle\) in Eqs. (36)--(38) is based on the
Lorentz-like transformation, in which \(\langle\beta_t\rangle\) is
also the average transverse velocity of the motion reference
system which reflects the collective flow. This treatment is
different from the blast-wave model~\cite{47,48,49,50}, though the
results are not contradictory.

Although there are many works already done on analysis and on the
interpretation of the Tsallis statistics role in high energy
collisions~\cite{32b,32c,32d,32e,32f}, the improvement of the
present work is significant. 1) The revised index \(a_0\) which
flexibly describes the winding degree of the spectra in
low-\(p_T\) region is introduced, in which the contribution of
resonance generation is significant. 2) The revised Tsallis-like
function is used to describe the transverse momenta of the
participant partons which contribute to \(p_T\) of particles,
where various possible azimuths in the transverse plane are
discussed. 3) The average transverse flow velocity is introduced,
and the kinetic freeze-out temperature and average transverse flow
velocity are conveniently obtained at the parton level.
\\

\section{Summary and conclusion}

To see the method for describing the transverse momentum spectra
of final-state particles produced in high energy collisions, the
physics picture and formalism expression treated in the framework
of multi-source thermal model at the parton level has been
reviewed. Generally, two or three partons contribute to the
transverse momentum spectra of mesons or baryons, while two
partons contribute to the transverse momentum spectra of leptons
or jets.

In general case, the difference between the parton azimuths is
variant in \([0,2\pi]\). The Monte Carlo method may be used to
perform the calculations. If the difference is 0 or $\pi$, one may
obtain a convolution of two or three probability density
functions, which is an analytical expression. If the difference is
\(\pi/2\), one may also obtain an analytical expression if one
integrates azimuthal variable over the united probability density
function of transverse momentum and azimuth.

The fitting experience shows that the convolution of two or three
probability density functions is more suitable in describing the
particle transverse momentum spectra, though the various
differences between the parton azimuths sounds more reasonable.
The transverse momentum spectra in different collisions are
uniformly described at the parton level. The same contributor
partons reflect the origin of the similarity, commonality, and
universality existed in high energy collisions.
\\
\\
\\
{\bf Data Availability}

The data used to support the findings of this study are included
within the article and are cited at relevant places within the
text as references.
\\
\\
{\bf Ethical Approval}

The authors declare that they are in compliance with ethical
standards regarding the content of this paper.
\\
\\
{\bf Disclosure}

The funding agencies have no role in the design of the study; in
the collection, analysis, or interpretation of the data; in the
writing of the manuscript; or in the decision to publish the
results.
\\
\\
{\bf Conflicts of Interest}

The authors declare that there are no conflicts of interest
regarding the publication of this paper.
\\
\\
{\bf Acknowledgments}

This work was supported by the National Natural Science Foundation
of China under Grant Nos. 12147215, 12047571, 11575103, and
11947418, the Scientific and Technological Innovation Programs of
Higher Education Institutions in Shanxi (STIP) under Grant No.
201802017, the Shanxi Provincial Natural Science Foundation under
Grant No. 201901D111043, and the Fund for Shanxi ``1331 Project"
Key Subjects Construction.


\begin{thebibliography}{99}

\bibitem{1}
H. Wang, J.-H. Chen, Y.-G. Ma, and S. Zhang, ``Charm hadron
azimuthal angular correlations in Au+Au collisions at
$\sqrt{s_{NN}}=200$ GeV from parton scatterings," {\it Nuclear
Science and Techniques}, vol. 30, no. 12, article 185, 2019.

\bibitem{2}
Z.-B. Tang, W.-M. Zha, and Y.-F. Zhang, ``An experimental review
of open heavy flavor and quarkonium production at RHIC," {\it
Nuclear Science and Techniques,} vol. 31, no. 8, article 81, 2020.

\bibitem{3}
Y.-C. Liu and X.-G. Huang, ``Anomalous chiral transports and spin
polarization in heavy-ion collisions," {\it Nuclear Science and
Techniques}, vol. 31, no. 6, article 56, 2020.

\bibitem{4}
E. K. G. Sarkisyan and A. S. Sakharov, ``Multihadron production
features in different reactions," Invited talk at the XXXV
International Symposium on Multiparticle Dynamics (ISMD 05),
Kromeriz, Czech Republic, 9--15 Aug. 2005, {\it AIP Conference
Proceedings}, vol. 828, pp. 35--41, 2006.

\bibitem{5}
E. K. G. Sarkisyan and A. S. Sakharov, ``Relating multihadron
production in hadronic and nuclear collisions," {\it The European
Physical Journal C}, vol. 70, no. 3, pp. 533--541, 2010.

\bibitem{6}
A. N. Mishra, R. Sahoo, E. K. G. Sarkisyan, and A. S. Sakharov,
``Effective-energy budget in multiparticle production in nuclear
collisions," {\it The European Physical Journal C}, vol. 74, no.
11, article 3147, 2014 and ``Erratum," {\it ibid}, vol. 75, no. 2,
article 70, 2015.

\bibitem{7}
E. K. G. Sarkisyan, A. N. Mishra, R. Sahoo, and A. S. Sakharov,
``Multihadron production dynamics exploring the energy balance in
hadronic and nuclear collisions," {\it Physical Review D}, vol.
93, no. 5, article 054046, 2016 and ``Erratum," {\it ibid}, vol.
93, no. 7, article 079904, 2016.

\bibitem{8}
E. K. G. Sarkisyan, A. N. Mishra, R. Sahoo, and A. S. Sakharov,
``Centrality dependence of midrapidity density from GeV to TeV
heavy-ion collisions in the effective-energy universality picture
of hadroproduction," {\it Physical Review D}, vol. 94, no. 1,
article 011501(R), 2016.

\bibitem{9}
E. K. G. Sarkisyan, A. N. Mishra, R. Sahoo, and A. S. Sakharov,
``Effective-energy universality approach describing total
multiplicity centrality dependence in heavy-ion collisions," {\it
EPL (Europhysics Letters)}, vol. 127, no. 6, article 62001, 2019.

\bibitem{10}
A. N. Mishra, A. Ortiz, and G. Paic, ``Intriguing similarities of
high-$p_T$ particle production between $pp$ and $A$-$A$
collisions," {\it Physical Review C}, vol. 99, no. 3, article
034911, 2019.

\bibitem{11}
P. Castorina, A. Iorio, D. Lanteri, H. Satz, and M. Spousta,
``Universality in hadronic and nuclear collisions at high energy,"
{\it Physical Review C}, vol. 101, no. 5, article 054902, 2020.

\bibitem{12}
F.-H. Liu, ``Unified description of multiplicity distributions of
final-state particles produced in collisions at high energies,"
{\it Nuclear Physics A}, vol. 810, nos. 1--4, pp. 159--172, 2008.

\bibitem{13}
F.-H. Liu and J.-S. Li, ``Isotopic production cross section of
fragments in $^{56}$Fe+p and $^{136}$Xe($^{124}$Xe)+Pb reactions
over an energy range from 300A to 1500A MeV," {\it Physical Review
C}, vol. 78, no. 4, article 044602, 2008.

\bibitem{14}
F.-H. Liu, ``Dependence of charged particle pseudorapidity
distributions on centrality and energy in $p(d)A$ collisions at
high energies," {\it Physical Review C}, vol. 78, no. 1, article
014902, 2008.

\bibitem{15}
F.-H. Liu, B. K. Singh, and N. N. Abd Allah, ``Integral frequency
distribution of projectile alpha fragments in nuclear
multifragmentations at intermediate and high energies," {\it
Nuclear Physics B (Proceedings Supplements)}, vols. 175--176, pp.
54--57, 2008.

\bibitem{16}
F.-H. Liu, Q.-W. L{\"u}, B.-C. Li, and R. Bekmirzaev, ``A
description of the multiplicity distributions of nuclear fragments
in $hA$ and $AA$ collisions at intermediate and high energies,"
{\it Chinese Journal of Physics}, vol. 49, no. 2, pp. 601--620,
2011.

\bibitem{17}
W. Y. Chang, ``Jets induced in emulsion and cloud chambers by
cosmic ray particles of energy ($10^{11}$--$10^{14}$ eV)," {\it
Acta Physica Sinica}, vol. 17, no. 8, pp. 9--33, 1961.

\bibitem{18}
G. D. Westfall, J. Gosset, P. J. Johansen, A. M. Poskanzer, W. G.
Meyer, H. H. Gutbrod, A. Sandoval, and R. Stock, ``Nuclear
fireball model for proton inclusive spectra from relativistic
heavy-ion collisions," {\it Physical Review Letters}, vol. 37, no.
18, pp. 1202--1205, 1976.

\bibitem{19}
G. Ingrosso and P. Rotelli, ``Three-fireball model of $\pi^-p$
inelastic interactions at 205 GeV/c," {\it Il Nuovo Cimento A},
vol. 41, no. 2, pp. 233--249, 1977.

\bibitem{20}
A. D'Innocenzo, G. Ingrosso, and P. Rotelli, ``The $p\bar p$
annihilation channel as a prototype for the central firball in
$pp$ production processes," {\it Lettere al Nuovo Cimento}, vol.
25, no. 13, pp. 393--398, 1979.

\bibitem{21}
A. D'Innocenzo, G. Ingrosso, and P. Rotelli, ``The three-component
fireball model and $pp$ interactions," {\it Lettere al Nuovo
Cimento}, vol. 27, no. 14, pp. 457--466, 1980.

\bibitem{22}
A. D'Innocenzo, G. Ingrosso, and P. Rotelli, ``A universal scaling
function for hadron-hadron interactions," {\it Il Nuovo Cimento
A}, vol. 55, no. 4, pp. 417--436, 1980.

\bibitem{23}
K.-C. Chou, L.-S. Liu, and T.-C. Meng, ``Koba-Nielsen-Olesen
scaling and production mechanism in high-energy collisions," {\it
Physical Review D}, vol. 28, no. 5, pp. 1080--1085, 1983.

\bibitem{24}
L.-S. Liu and T.-C. Meng, ``Multiplicity and energy distributions
in high-energy $e^+e^-$, $pp$, and $p\bar p$ collisions," {\it
Physical Review D}, vol. 27, no. 11, pp. 2640--2647, 1983.

\bibitem{25}
F.-H. Liu, J.-S. Li, and M.-Y. Duan, ``Light fragment emission in
$^{86}$Kr-$^{124}$Sn collisions at 25 MeV/nucleon," {\it Physical
Review C}, vol. 75, no. 5, article 054613, 2007.

\bibitem{26}
F.-H. Liu, ``Particle pseudorapidity distribution in Au-Au
collisions at $\sqrt{s}=130A$ GeV," {\it Physical Review C}, vol.
66, no. 4, article 047902, 2002.

\bibitem{27}
F.-H. Liu, ``Particle production in Au-Au collisions at RHIC
energies," {\it Physics Letters B}, vol. 583, nos. 1--2, pp.
68--72, 2004.

\bibitem{28}
F.-H. Liu, Y.-Q. Gao, T. Tian, and B.-C. Li, ``Unified description
of transverse momentum spectrums contributed by soft and hard
processes in high-energy nuclear collisions," {\it The European
Physical Journal A}, vol. 50, no. 6, article 94, 2014.

\bibitem{29}
P.-P. Yang, F.-H. Liu, and R. Sahoo, ``A new description of
transverse momentum spectra of identified particles produced in
proton-proton collisions at high energies," {\it Advances in High
Energy Physics}, vol. 2020, article 6742578, 2020.

\bibitem{30}
P.-P. Yang, M.-Y. Duan, and F.-H. Liu, ``Dependence of related
parameters on centrality and mass in a new treatment for
transverse momentum spectra in high energy collisions," {\it The
European Physical Journal A}, vol. 57, no. 2, article 63, 2021.

\bibitem{31}
L.-L. Li, F.-H. Liu, and Kh. K. Olimov, ``Excitation functions of
Tsallis-like parameters in high-energy nucleus-nucleus
collisions," {\it Entropy}, vol. 23, no. 4, article 478, 2021.

\bibitem{31a}
F.-H. Liu, Y.-Q. Gao, H.-R. Wei, ``On descriptions of particle
transverse momentum spectra in high energy collisions," {\it
Advances in High Energy Physics}, vol. 2014, article 293873, 2014.

\bibitem{31a0}
L.-L. Li and F.-H. Liu, ``Kinetic freeze-out properties from
transverse momentum spectra of pions in high energy proton-proton
collisions," {\it Physics}, vol. 2, no. 2, pp. 277--308, 2020.

\bibitem{31a1}
T. Chujo, ``Excitation functions of baryon anomaly and freeze-out
properties at RHIC-PHENIX," {\it Journal of Physics G}, vol. 34,
no. 8, pp. S893--S896, 2007.

\bibitem{31a2}
H. Petersen, J. Steinheimer, M. Bleicher, and H. Stoecker,
``$\langle m_T\rangle$ excitation function: Freeze-out and
equation of state dependence," {\it Journal of Physics G}, vol.
36, no. 5, article 055104, 2009.

\bibitem{31a3}
J. Cleymans, H. Oeschler, K. Redlich, and S. Wheaton,
``Strangeness excitation functions and transition from baryonic to
mesonic freeze-out," {\it Acta Physica Polonica B Proceedings
Series}, vol. 3, no. 3, pp. 533--538, 2010.

\bibitem{31a4}
Y. Nara and H. Stoecker, ``Sensitivity of the excitation functions
of collective flow to relativistic scalar and vector meson
interactions in the relativistic quantum molecular dynamics model
RQMD.RMF," {\it Physical Review C}, vol. 100, no. 5, article
054902, 2019.

\bibitem{31a5}
V. Kolesnikov, V. Kireyeu, V. Lenivenko, A. Mudrokh, K. Shtejer,
D. Zinchenko, and E. Bratkovskaya, ``A new review of excitation
functions of hadron production in pp collisions in the NICA energy
range," {\it Physics of Particles and Nuclei Letters}, volume 17,
no. 2, pp. 142--153, 2020.

\bibitem{31a7}
I. Arsene, I. G. Bearden, D. Beavisa et al. (BRAHMS
Collaboration), ``Quark-gluon plasma and color glass condensate at
RHIC? The perspective from the BRAHMS experiment," {\it Nuclear
Physics A}, vol. 757, nos. 1--2, pp. 1--27, 2005.

\bibitem{31a8}
B. B. Back, M. D. Baker, M. Ballintijn et al. (PHOBOS
Collaboration), ``The PHOBOS perspective on discoveries at RHIC,"
{\it Nuclear Physics A}, vol. 757, nos. 1--2, pp. 28--101, 2005.

\bibitem{31a9}
J. Adams, M. M. Aggarwal, Z. Ahammed et al. (STAR Collaboration),
``Experimental and theoretical challenges in the search for the
quark-gluon plasma: The STAR Collaboration's critical assessment
of the evidence from RHIC collisions," {\it Nuclear Physics A},
vol. 757, nos. 1--2, pp. 102--183, 2005.

\bibitem{31a10}
K. Adcox, S. S. Adler, and S. Afanasiev et al. (PHENIX
Collaboration), ``Formation of dense partonic matter in
relativistic nucleus-nucleus collisions at RHIC: Experimental
evaluation by the PHENIX Collaboration," {\it Nuclear Physics A},
vol. 757, nos. 1--2, pp. 184--283, 2005.

\bibitem{31a11}
K. Zapp, G. Ingelman, J. Rathsman, and J. Stachel, ``Heavy quark
energy loss through soft QCD scattering in the QGP," {\it
International Journal of Modern Physics E}, vol. 16, no. 8, pp.
2072--2078, 2007.

\bibitem{31a12}
I. Kuznetsova and J. Rafelski, ``Non-equilibrium heavy flavored
hadron yields from chemical equilibrium strangeness-rich QGP,"
{\it Journal of Physics G}, vol. 35, no. 4, article 044043, 2008.

\bibitem{31a13}
Z.-W. Lin, H. L. Li, and F. Q. Wang, ``Heavy quark flow as better
probes of QGP properties," {\it EPJ Web of Conferences}, vol. 171,
article 19005, 2018.

\bibitem{31a14}
B. V. Jacak, ``Quark matter: Status and challenges", {\it Nulcear
Physics A}, vol. 1005, article 122052, 2021.

\bibitem{31a15}
C. Shen, ``Studying QGP with flow: A theory overview," {\it
Nulcear Physics A}, vol. 1005, article 121788, 2021.

\bibitem{31a16}
K. K. Gajdosov{\'a}, ``Probing QGP with flow: An experimental
overview," {\it Nulcear Physics A}, vol. 1005, article 121802,
2021.

\bibitem{31a17}
K. J. Eskola, ``Pre-thermalization dynamics: initial conditions
for QGP at the LHC and RHIC from perturbative QCD," {\it Progress
of Theoretical Physics Supplement}, vol. 129, pp. 1--10, 1997.

\bibitem{31a18}
K. J. Eskola, ``Initial state of the QGP from perturbative QCD +
saturation," {\it Nuclear Physics A}, vol. 702, nos. 1--4, pp.
249--258, 2002.

\bibitem{31a19}
J. Letessier and J. Rafelski, ``QCD equations of state and the QGP
liquid model," {\it Physical Review C}, vol. 67, no. 3, article
031902, 2003.

\bibitem{31a20}
H. Satz, ``Critical behaviour in statistical QCD," {\it
International Journal of Modern Physics A}, vol. 21, no. 4, pp.
672--681, 2006.

\bibitem{31a21}
L. S. Kisslinger, ``Review of QCD, QGP, heavy quark meson
production enhancement and suppression," {\it International
Journal of Modern Physics A}, vol. 32, no. 15, article 1730008,
2017.

\bibitem{31aa1}
G. Q. Li, C. M. Ko, and G. E. Brown, ``Effects of in-medium vector
meson masses on low-mass dileptons from SPS heavy-ion collisions,"
{\it Nuclear Physics A}, vol. 606, nos. 3--4, pp. 568--606, 1996.

\bibitem{31aa2}
G. Q. Li, C. M. Ko, G. E. Brown, and H. Sorge, ``Dilepton
production in proton-nucleus and nucleus-nucleus collisions at SPS
energies," {\it Nuclear Physics A}, vol. 611, no. 4, pp. 539--567,
1996.

\bibitem{31aa3}
A. Bieniek, ``Modification of the $\pi$-$\omega$-$\rho$ vertex in
nuclear medium and its influence on the dilepton production rate
in relativistic heavy-ion collisions," arXiv:nucl-th/0411084,
2004.

\bibitem{31aa4}
B.-W. Zhang, C. M. Ko, and W. Liu, ``Thermal charm production in
quark-gluon plasma at LHC," {\it Physical Review C}, vol. 77, no.
2, article 024901, 2008.

\bibitem{31b}
Z.-J. Xiao and C.-D. L{\"u}, {\it Introduction to Particle
Physics}, Science Press, Beijing, China, 2016.

\bibitem{31c}
P. Desgrolard, M. Giffon, and E. Martynov, ``Elastic pp and
$\overline{p}p$ scattering in the Modified Additive Quark Model,"
{\it The European Physical Journal C}, vol. 18, no. 12, pp.
359--367, 2000.

\bibitem{31d}
Yu. M. Shabelski and A. G. Shuvaev, ``Midrapidity inclusive
densities in high energy $pp$ collisions in additive quark model,"
{\it The European Physical Journal C}, vol. 76, no. 8, article
470, 2016.

\bibitem{31e}
Yu. M. Shabelski and A. G. Shuvaev, ``Real part of $pp$ scattering
amplitude in Additive Quark Model at LHC energies," {\it The
European Physical Journal C}, vol. 78, no. 6, article 497, 2018.

\bibitem{31f}
G. H. Arakelyan, Yu. M. Shabelski, and A. G. Shuvaev, ``Central
and peripheral hadron-nucleus collisions in the Additive Quark
Model," arXiv:1805.11293 [hep-ph], 2018.

\bibitem{32}
P.-P. Yang, Q. Wang, and F.-H. Liu, ``Mutual derivation between
arbitrary distribution forms of momenta and momentum components,"
{\it International Journal of Theoretical Physics}, vol. 58, no.
8, pp. 2603--2618, 2019.

\bibitem{32a}
G.-R. Zhou, {\it Probability Theory and Mathematical Statsitics},
Higher Education Press, Beijing, China, 1984.

\bibitem{32b}
C. Tsallis, ``Possible generalization of Boltzmann-Gibbs
statistics," {\it Journal of Statistical Physics}, vol. 52, nos.
1--2, pp. 479--487, 1988.

\bibitem{32c}
T. S. Bir{\'o}, G. Purcsel, K. {\"U}rm{\"o}ssy, ``Non-extensive
approach to quark matter," {\it The European Physical Journal A},
vol. 40, no. 3, pp. 325--340, 2009.

\bibitem{32d}
J. Cleymans and D. Worku, ``Relativistic thermodynamics:
Transverse momentum distributions in high-energy physics," {\it
The European Physical Journal A}, vol. 48, no. 11, article 160,
2012.

\bibitem{32e}
H. Zheng and L. L. Zhu, ``Comparing the Tsallis distribution with
and without thermodynamical description in p+p collisions," {\it
Advances in High Energy Physics}, vol. 2016, article 9632126,
2016.

\bibitem{32f}
J. Cleymans and M. W. Paradza, ``Tsallis statistics in high energy
physics: Chemical and thermal freeze-outs," {\it Physics}, vol. 2,
no. 4, pp. 654--664, 2020.

\bibitem{45}
Y.-M. Tai, P.-P. Yang, F.-H. Liu, ``An analysis of transverse
momentum spectra of various jets produced in high energy
collisions," {\it Advances in High Energy Physics}, vol. 2021,
article 8832892, 2021.

\bibitem{33a}
S. Takeuchi, K. Murase, T. Hirano et al., ``Effects of hadronic
rescattering on multistrange hadrons in high-energy nuclear
collisions," {\it Physical Review C}, vol. 92, no. 4, article
044907, 2015.

\bibitem{33b}
H. Heiselberg and A. M. Levy, ``Elliptic flow and
Hanbury-Brown-Twiss in noncentral nuclear collisions," {\it
Physical Review C}, vol. 59, no. 5, pp. 2716--2727, 1999.

\bibitem{33c}
U. W. Heinz, ``Concepts of heavy-ion physics," Lecture Notes for
Lectures Presented at the 2nd CERN-Latin-American School of
High-Energy Physics, June 1--14, 2003 (San Miguel Regla, Mexico,
2004), arXiv:hep-ph/0407360, 2004.

\bibitem{33d}
R. Russo, ``Measurement of $D^+$ meson production in p-Pb
collisions with the ALICE detector," Ph.D. thesis (Universita
degli Studi di Torino, Italy, 2015), arXiv:1511.04380 [nucl-ex],
2015.

\bibitem{33}
H.-R. Wei, F.-H. Liu, and R. A. Lacey, ``Kinetic freeze-out
temperature and flow velocity extracted from transverse momentum
spectra of final-state light flavor particles produced in
collisions at RHIC and LHC," {\it The European Physical Journal
A}, vol. 52, no. 4, article 102, 2016.

\bibitem{34}
H.-R. Wei, F.-H. Liu, and R. A. Lacey, ``Disentangling random
thermal motion of particles and collective expansion of source
from transverse momentum spectra in high energy collisions," {\it
Journal of Physics G}, vol. 43, no. 12, article 125102, 2016.

\bibitem{35}
Kh. K. Olimov, A. Iqbal, and S. Masood, ``Systematic analysis of
midrapidity transverse momentum spectra of identified charged
particles in $p$+$p$ collisions at $(s_{nn})^{1/2}=2.76$, 5.02,
and 7 TeV at the LHC," {\it International Journal of Modern
Physics A}, vol. 35, no. 27, article 2050167, 2020.

\bibitem{36}
Kh. K. Olimov, S. Z. Kanokova, K. Olimov, K. G. Gulamov, B. S.
Yuldashev, S. L. Lutpullaev, and F. Y. Umarov, ``Average
transverse expansion velocities and global freeze-out temperatures
in central Cu+Cu, Au+Au, and Pb+Pb collisions at high energies at
RHIC and LHC," {\it Modern Physics Letters A}, vol. 35, no. 14,
article 2050115, 2020.

\bibitem{37}
Kh. K. Olimov, S. Z. Kanokova, A. K. Olimov, K. I. Umarov, B. J.
Tukhtaev, K. G. Gulamov, B. S. Yuldashev, S. L. Lutpullaev, N. Sh.
Saidkhanov, K. Olimov, and T. Kh. Sadykov, ``Combined analysis of
midrapidity transverse momentum spectra of the charged pions and
kaons, protons and antiprotons in $p$+$p$ and Pb+Pb collisions at
$(s_{nn})^{1/2}=2.76$ and 5.02 TeV at the LHC," {\it Modern
Physics Letters A}, vol. 35, no. 29, article 2050237, 2020.

\bibitem{38}
Kh. K. Olimov, K. I. Umarov1, A. Iqbal, S. Masood, and F.-H. Liu,
``Analysis of midrapidity transverse momentum distributions of the
charegd pions and kaons, protons and antiprotons in $p$+$p$
collisions at $(s_{nn})^{1/2}=2.76$, 5.02, and 7 TeV at the LHC,"
{\it Proceedings of the International Conference on ``Fundamental
and Applied Problems of Physics"}, pp. 78--83, Tashkent,
Uzbekistan, September 22--23, 2020.

\bibitem{46a}
A. Adare et al. (PHENIX Collaboration), ``Measurement of neutral
mesons in p+p collisions at $\sqrt{s}=200$ GeV and scaling
production," {\it Physical Review D}, vol. 83, no. 5, article
052004, 2011.

\bibitem{47}
E. Schnedermann, J. Sollfrank, and U. Heinz, ``Thermal
phenomenology of hadrons from 200A GeV S+S collisions," {\it
Physical Review C}, vol. 48, no. 5, pp. 2462--2475, 1993.

\bibitem{48}
B. I. Abelev et al. (STAR Collaboration), ``Systematic
measurements of identified particle spectra in pp, d+Au, and Au+Au
collisions at the STAR detector," {\it Physical Review C}, vol.
79, no. 3, article 034909, 2009.

\bibitem{49}
Z. B. Tang, Y. C. Xu, L. J. Ruan, G. van Buren, F. Q. Wang, and Z.
B. Xu, ``Spectra and radial flow in relativistic heavy ion
collisions with Tsallis statistics in a blastwave description,"
{\it Physical Review C}, vol. 79, no. 5, article 051901(R), 2009.

\bibitem{50}
B. I. Abelev et al. (STAR Collaboration), ``Identified particle
production, azimuthal anisotropy, and interferometry measurements
in Au+Au collisions at $\sqrt{s_{NN}}=9.2$ GeV," {\it Physical
Review C}, vol. 81, no. 2, article 024911, 2010.

\end{thebibliography}
\end{document}